\documentclass[aps,prl,superscriptaddress,twocolumn,longbibliography]{revtex4-2}
\usepackage{bbm}
\usepackage{graphicx}
\usepackage{dcolumn}
\usepackage{bm}
\usepackage{subfigure}
\usepackage{amsmath}
\usepackage{amsfonts}
\usepackage{feynmf}
\usepackage{hyperref}

\usepackage{attachfile}

\newcommand{\bk}{\boldsymbol k}

\newcommand{\zb}{\color {black}}

\usepackage{times}

\begin{document}
\title{Coexistence of Chiral Majorana Edge States and Bogoliubov Fermi Surfaces  in Two-Dimensional\\
 Nonsymmorphic Dirac Semimetal/Superconductor Heterostructures}

\author{Yijie Mo}
\altaffiliation{These authors contributed equally to this work.}
\affiliation{Guangdong Provincial Key Laboratory of Magnetoelectric Physics and Devices, State Key Laboratory of Optoelectronic Materials and Technologies, School of Physics, Sun Yat-sen University, Guangzhou 510275, China}

\author{Xiao-Jiao Wang}
\altaffiliation{These authors contributed equally to this work.}
\affiliation{Guangdong Provincial Key Laboratory of Magnetoelectric Physics and Devices, State Key Laboratory of Optoelectronic Materials and Technologies, School of Physics, Sun Yat-sen University, Guangzhou 510275, China}

\author{Zheng-Yang Zhuang}
\affiliation{Guangdong Provincial Key Laboratory of Magnetoelectric Physics and Devices, State Key Laboratory of Optoelectronic Materials and Technologies, School of Physics, Sun Yat-sen University, Guangzhou 510275, China}

\author{Zhongbo Yan}
\email{yanzhb5@mail.sysu.edu.cn}
\affiliation{Guangdong Provincial Key Laboratory of Magnetoelectric Physics and Devices, State Key Laboratory of Optoelectronic Materials and Technologies, School of Physics, Sun Yat-sen University, Guangzhou 510275, China}

\date{\today}

\begin{abstract}
Dirac semimetals are renowned for the host of singular symmetry-protected band degeneracies which can give rise to
other exotic phases. In this work, we consider a two-dimensional Dirac semimetal 
stabilized by $\mathcal{PT}$ symmetry and nonsymmorphic symmetries. We find that an out-of-plane Zeeman field
can lift the Dirac points and transform the system into a Chern insulator with chiral edge states.
By placing the nonsymmorphic Dirac semimetal in proximity
to an $s$-wave superconductor, we uncover that chiral topological superconductors with large Chern numbers can be achieved. In addition,
we find that topologically-protected Bogoliubov Fermi surfaces can also emerge in this system, due to
the coexistence of inversion symmetry and particle-hole symmetry. Notably, we find that the chiral Majorana edge state persists even when the Chern number becomes ill-defined due to the appearance of Bogoliubov Fermi surfaces. The impact of these Bogoliubov Fermi surfaces on the thermal Hall effects is also investigated. Our study not only identifies a class of materials capable of realizing topological
Bogoliubov Fermi surfaces through conventional $s$-wave superconductivity, but also uncovers an exotic phase
where chiral Majorana edge states and Bogoliubov Fermi surfaces coexist.
\end{abstract}

\maketitle

In the past two decades, chiral topological superconductors (CTSCs)
have garnered significant attention due to the host of many exotic properties~\cite{qi2011,alicea2012new,leijnse2012introduction,Tanaka2012,stanescu2013majorana,Beenakker2013,Elliott2015,Sato2016jpsj,Kallin2016review}.
CTSCs refer to two-dimensional (2D) fully-gapped superconductors
characterized by a nonzero Chern number. Their defining characteristics include
the presence of robust chiral Majorana edge states and quantized thermal Hall effects
in the low-temperature limit~\cite{read2000,Nomura2012}. Additionally, when the Chern number is odd,
vortices in the CTSCs will harbor robust Majorana zero modes
which exhibit non-Abelian
statistics~\cite{ivanov2001} and hold potential for applications in topological quantum computation~\cite{nayak2008review,sarma2015majorana,Karzig2017MZM,Marra2022}.
While CTSCs can naturally arise from pairings with nonzero angular momentum, such
as $p+ip$, $d+id$ and $f+if$ pairings~\cite{Schaffer2012CTSC,Liu2013CTSC,Qin2019CTSC}, the scarcity of such pairings in superconducting
materials has prompted the exploration of alternative pathways.
A significant breakthrough in theory is the recognition that the combination
of antisymmetric spin-orbit coupling (SOC), Zeeman field with direction perpendicular
to the spin plane of the SOC, and conventional $s$-wave pairing—all prevalent
in real materials—can lead to the realization of CTSCs~\cite{fu2008,Fujimoto2008TSC,zhang2008px,sato2009non,sau2010,alicea2010,qi2010chiral,Li2016shiba}.
This theoretical advancement has sparked experimental efforts to pursue CTSCs,
and notable progress has been witnessed in this endeavor~\cite{Menard2017,Alexandra2019,Kezilebieke2020,Li2024observation}.

Recently, Bogoliubov Fermi surfaces (BFSs) in superconductors have generated
considerable interest due to their profound impact on various transport properties of these materials~\cite{Brydon2018BFS,Lapp2020BFS,Setty2020NC,Setty2020BFS,Herbut2021BFS,Timm2021BFS,Oh2021BFS,Jiang2021BFS,Banerjee2022BFS,Cao2023BFS,
Serafim2024BFS,Mateos2024BFS,Pal2024BFS,Min2024BFS,Ohashi2024,Miki2024BFS,Pal2024BFS}.
BFSs represent the Fermi surfaces of Bogoliubov quasiparticles, signifying a finite density of states for zero-energy excitations~\cite{volovik1989}. Typically, the presence of zero-energy excitations is detrimental to superconductors, as it suggests that breaking Cooper pairs incurs no energy cost. However, recent theoretical studies have demonstrated that BFSs can be stabilized under specific conditions~\cite{Menke2019BFS,Link2020BFS,Autti2020BFS,Oh2020BFS,Bhattacharya2023BFS}. Intriguingly, while a global topological invariant defined over the Brillouin
zone is no longer viable for superconductors with BFSs, the BFS itself can possess topological protection~\cite{Sumita2019classify,Tomas2017}. In a seminal paper, Agterberg {\it et al.} revealed that multiband systems with intrinsic
time-reversal-symmetry-breaking interband pairings can exhibit BFSs protected by a $Z_{2}$ invariant
if the inversion symmetry is preserved~\cite{Agterberg2017BFS}. In the study of pairing in 3D spin-orbit coupled
spin-$3/2$ semimetals, it is also discovered that some unconventional pairings can lead to BFSs~\cite{Timm2017BFS,Venderbos2018,Link2020RSW,Dutta2021BFS,Kobayashi2022BFS}.
Besides BFSs induced by pairings breaking time-reversal symmetry intrinsically, Yuan and Fu also revealed that
the combination of antisymmetric SOC, Zeeman field with direction lying
in the spin plane of the SOC, and conventional $s$-wave pairing
can also give rise to BFSs~\cite{Yuan2018BFS}.  This scenario was later demonstrated
in superconductor-topological insulator~\cite{Zhu2021BFS}
and superconductor-semiconductor hybrid systems~\cite{Phan2022}.

\begin{figure}[t]
\centering
\includegraphics[width=0.45\textwidth]{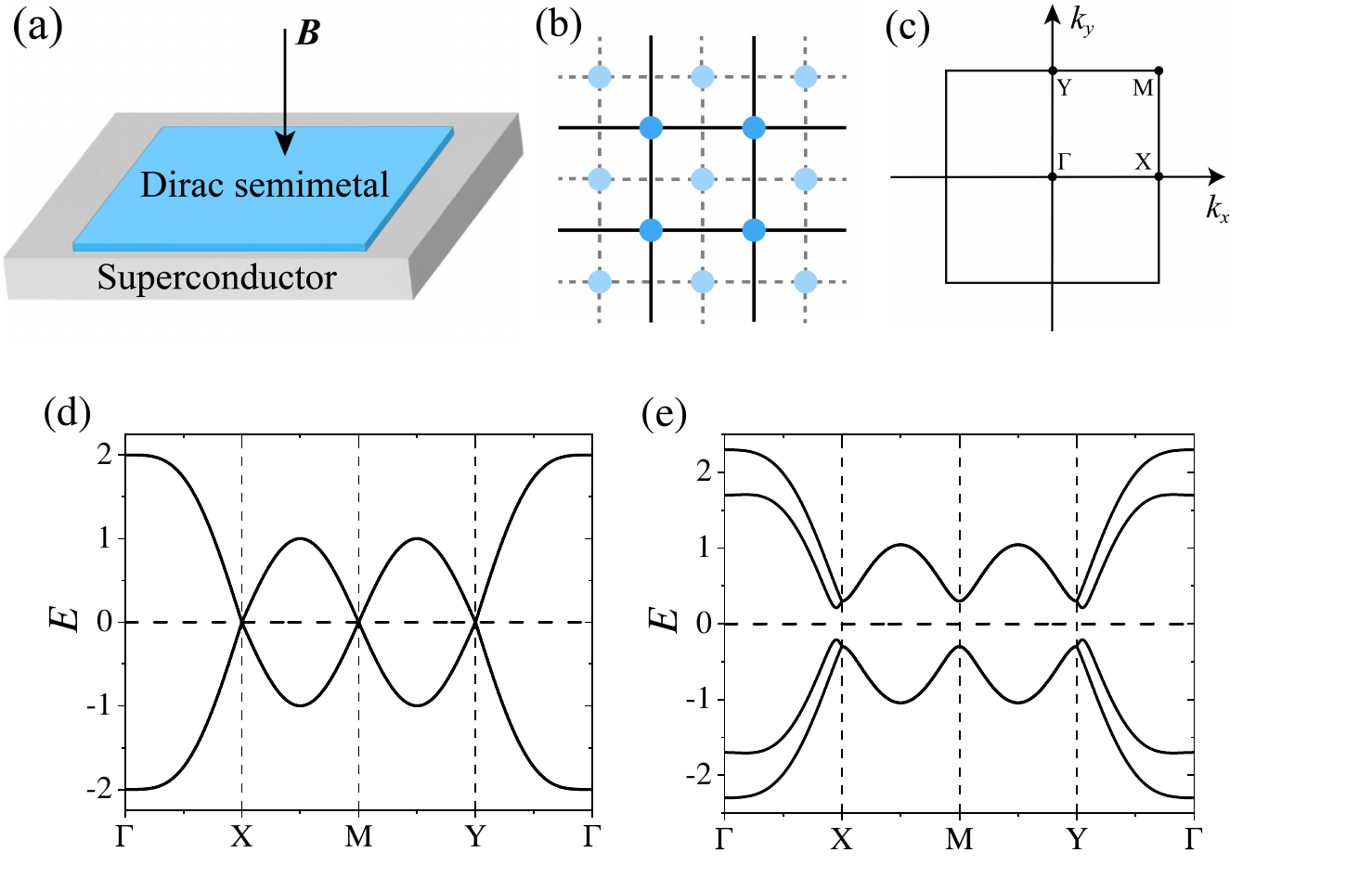}
\caption{(a) Schematic of the heterostructure. (b) Lattice structure
of the DSM. (c) Brillouin zone with the four time-reversal invariant
momentums highlighted. (d) Band structure along high symmetry lines within the Brillouin zone. The
parameter values are: $t=0$, $\lambda=0.5$, $\eta=0.5$ and $M_{z}=0$. 
(e) Band structure with the Zeeman field set to $M_{z}=0.3$.
}\label{fig1}
\end{figure}

Interestingly, the combination of antisymmetric SOC, Zeeman field, and conventional $s$-wave pairing can give rise to both CTSCs and BFSs in two dimensions. However, these two phenomena have distinct requirements: CTSCs necessitate a perpendicular Zeeman field~\cite{sato2009non,sau2010}, whereas BFSs require a parallel Zeeman field aligned with the spin plane of the SOC~\cite{Yuan2018BFS}. Furthermore, it is important to note that since the Chern number is undefined in a gapless superconductor, CTSCs and BFSs inherently represent two incompatible phenomena. Therefore, it appears plausible to conclude that BFSs and chiral Majorana edge states would typically exist exclusively under their respective conditions. In this work, we investigate a 2D nonsymmorphic spin-orbit-coupled Dirac semimetal (DSM) under the influence of a conventional $s$-wave superconductor and the effect of a perpendicular Zeeman field,
as illustrated in Fig.\ref{fig1}(a). Our findings demonstrate that, contrary to expectations, BFSs and chiral Majorana edge states can coexist within this system.

{\it 2D nonsymmorphic DSM and Zeeman-field-induced Chern insulator.---}In two dimensions,
stabilizing fourfold degenerate Dirac points in the presence of SOC requires not only
$\mathcal{PT}$ symmetry, but also nonsymmorphic symmetries~\cite{Young2015DSM}. In this
work, we consider a DSM realized on the bipartite lattice as illustrated in Fig.\ref{fig1}(b).
The tight-binding Hamiltonian describing the DSM is given by~\cite{Young2015DSM}
\begin{eqnarray}
\mathcal{H}_{\rm DSM}(\bk)=\epsilon_{0}(\bk)+\epsilon_{1}(\bk)\sigma_{x}+\Lambda_{\rm so}(\bk)\sigma_{z}\label{Hamiltonian}
\end{eqnarray}
under the basis $\psi^{\dag}_{\bk}=(c_{A,\uparrow,\bk}^{\dag},c_{A,\downarrow,\bk}^{\dag},
c_{B,\uparrow,\bk}^{\dag},c_{B,\downarrow,\bk}^{\dag})$. Here
$\epsilon_{0}(\bk)=-2t(\cos k_{x}+\cos k_{y})$ denotes the kinetic energy
arising from the nearest-neighbor intra-sublattice hopping,
$\epsilon_{1}(\bk)=-4\eta\cos\frac{k_{x}}{2}\cos\frac{k_{y}}{2}$ denotes
the kinetic energy originating from the nearest-neighbor inter-sublattice hopping,
and $\Lambda_{\rm so}(\bk)=2\lambda_{\rm so}(\sin k_{y}s_{x}-\sin k_{x}s_{y})$
is the SOC.  The Pauli matrices $\sigma_{i}$ and $s_{i}$ act on the sublattice ($A,B$) and
spin ($\uparrow, \downarrow$) degrees of freedom, respectively. For notational simplicity,
the lattice constants are set to unity throughout this work,  and identity
matrices in orbital and spin space are made implicit.

The Hamiltonian in Eq.(\ref{Hamiltonian}) has inversion symmetry
($\mathcal{P}=\sigma_{x}$) despite the presence of SOC. Such inversion-symmetric SOC can emerge when
the system locally breaks the inversion symmetry but preserves it globally~\cite{Zhang2014}. The Hamiltonian also
has  time-reversal symmetry
($\mathcal{T}=is_{y}\mathcal{K}$ with $\mathcal{K}$ the complex conjugate operator),
a glide mirror symmetry $\{\mathcal{M}_{z}|(\frac{1}{2},\frac{1}{2})\}$, and two screw symmetries $\{\mathcal{C}_{2x}|(\frac{1}{2},0)\}$
and $\{\mathcal{C}_{2y}|(0,\frac{1}{2})\}$\cite{Young2015DSM}. Because of the $\mathcal{P}\mathcal{T}$ symmetry and the three nonsymmorphic
symmetries, this Hamiltonian harbors three fourfold degenerate Dirac points at the boundary
of the Brillouin zone, with their positions located at the three time-reversal invariant momentums (TRIMs),
$\mathbf{X}=(\pi,0)$, $\mathbf{Y}=(0,\pi)$ and $\mathbf{M}=(\pi,\pi)$, as shown in Figs.\ref{fig1}(c) and \ref{fig1}(d).

\begin{figure}[t]
\centering
\includegraphics[width=0.45\textwidth]{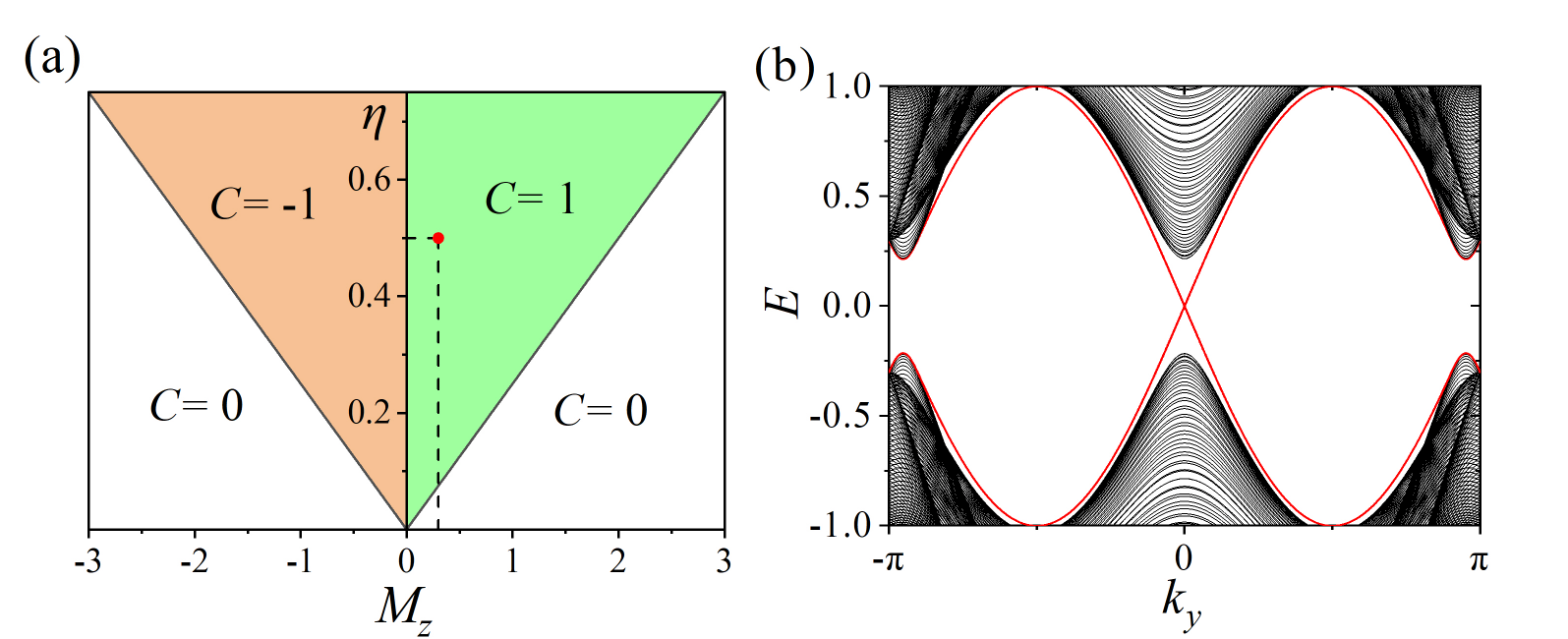}
\caption{(a) Phase diagram valid as long as $\lambda_{\rm so}\neq0$. (b) A Chern insulator with one branch of chiral edge state. 
The value of the shared parameters are: $t=0$, $\lambda_{\rm so}=0.5$, $\eta=0.5$ and $M_{z}=0.3$. 
}\label{fig2}
\end{figure}

It is known that breaking time-reversal symmetry can lift Dirac points, {\zb resulting in Chern insulators characterized by gapless chiral edge states~\cite{Haldane1988,yu2010a,chang2013experimental}. The most straightforward way to break
time-reversal symmetry is through the application of a spin-splitting Zeeman field.
However, in DSMs lacking SOC and thus exhibiting full spin-rotation symmetry (e.g., monolayer graphene), 
a Zeeman field cannot open a gap at the Dirac points. Instead, it shifts the energy bands with opposite spin polarizations in opposite directions, causing the fourfold spin-degenerate Dirac points to split into twofold spin-polarized Weyl points. In contrast, for the spin-orbit-coupled DSM described by the Hamiltonian in Eq.\ref{Hamiltonian},
we observe a markedly different behavior.  To demonstrate this,}
{\zb consider an out-of-plane Zeeman field oriented perpendicular to the spin plane of the SOC.} 
Accordingly,  the Hamiltonian becomes
\begin{eqnarray}
\mathcal{H}_{\rm CI}(\bk)=\epsilon_{0}(\bk)+\epsilon_{1}(\bk)\sigma_{x}+\Lambda_{\rm so}(\bk)\sigma_{z}+M_{z}s_{z},\label{ChernH}
\end{eqnarray}
where $M_{z}=g\mu_{B}B_{z}/2$ denotes the strength of Zeeman field induced, e.g., by
a $z$-directional magnetic field, i.e., $\mathbf{B}=B_{z}\hat{\mathbf{z}}$.
As the first term does not affect the band topology, in this part we let $\epsilon_{0}(\bk)=0$ {\zb by setting 
$t=0$} to simplify the discussion. The corresponding energy spectra are given by
\begin{eqnarray}
E_{\alpha,\beta}(\bk)=\alpha\sqrt{[\epsilon_{1}(\bk)+\beta M_{z}]^{2}+\Lambda^{2}(\bk)},
\end{eqnarray}
where $\Lambda(\bk)=2\lambda_{\rm so}\sqrt{\sin^{2}k_{x}+\sin^{2}k_{y}}$, $\alpha=\pm1$ and $\beta=\pm 1$.
It is easy to see that, once $M_{z}\neq0$,
the three Dirac points at $\mathbf{X}$, $\mathbf{Y}$ and $\mathbf{M}$ are all gapped,
as depicted in Fig.\ref{fig1}(e).  However, as $M_{z}$ reaches $\pm4\eta$,
the energy gap closes and reopens at $\boldsymbol{\Gamma}=(0,0)$,
indicating the occurrence of topological phase transitions.
Since the energy gap remains open whenever the absolute value of $M_{z}$ exceeds $4\eta$ (assuming $\eta$ is positive),
it is immediately evident that the system is topologically trivial in this regime.
Conversely, in the range of $0<|M_{z}|<4\eta$, the gapped phases are expected to be topological.
To confirm this, we numerically calculate the Chern number and determine the phase diagram shown in Fig.\ref{fig2}(a). Within the topological region, we find that the system with a cylindrical geometry possesses one gapless chiral edge state, which aligns with the calculated Chern number, as illustrated in Fig.\ref{fig2}(b).

{\it Nonsymmorphic DSM in proximity to an $s$-wave superconductor.---}The
SOC in the DSM  holds the promise of topological superconductivity if superconductivity is successfully induced into the DSM. In this paper, we explore the scenario where the DSM is grown on the surface of a conventional fully-gapped $s$-wave superconductor [see Fig.\ref{fig1}(a)].
Through the proximity effect, the DSM inherits $s$-wave pairing from the underlying superconductor.
{\zb To ensure a strong proximity effect, we reintroduce the term $\epsilon_{0}(\bk)$ and assume 
the presence of sizable Fermi surfaces in the normal state of the DSM. It is important to note that, 
although the upper and lower bands of the Hamiltonian (\ref{ChernH}) are separated by a gap induced 
by the Zeeman field, Fermi surfaces can either be generated through doping or intrinsically exist 
if the parameter $t$ is much larger than all other parameters. }

The resulting superconducting state of the DSM can be effectively described by a Bogoliubov-de Gennes (BdG) Hamiltonian of the form~\cite{Zhang2024}
\begin{eqnarray}
\mathcal{H}(\bk)&=&[\epsilon_{0}(\bk)-\mu]\tau_{z}+\epsilon_{1}(\bk)\tau_{z}\sigma_{x}
+M_{z}\tau_{z}s_{z}+\Delta_{s}\tau_{y}s_{y}\nonumber\\
&&+2\lambda_{\rm so}(\sin k_{x}\tau_{z}\sigma_{z}s_{y}-\sin k_{y}\sigma_{z}s_{x}),
\end{eqnarray}
where $\tau_{i}$ are Pauli matrices
acting on the particle-hole space, and $\Delta_{s}$ denotes the proximity-induced pairing amplitude. In principle,
$\Delta_{s}$ can be comparable to or much smaller than the pairing amplitude of
the underlying superconductor~\cite{Xu2014proximity}, depending on the quality of the interface between
the DSM and the superconductor.

In the absence of a Zeeman field, the BdG Hamiltonian belongs to the DIII class
according to the tenfold-way classification~\cite{schnyder2008,kitaev2009}. Due to the momentum-independent
nature of $s$-wave pairing, the superconducting phase is always topologically trivial
in this case. However, upon breaking time-reversal symmetry through the application of a Zeeman field,
the Hamiltonian transitions to the D class, thereby enabling the existence of CTSCs~\cite{Zhang2024}.
Regardless of the presence or absence of the Zeeman field, the BdG Hamiltonian possesses
both inversion symmetry and particle-hole symmetry ($\Xi=\tau_{x}\mathcal{K}$),
resulting in a symmetric energy spectrum centered at $E=0$.
These two symmetries are crucial for the topological protection of the BFSs discussed later.

To demonstrate that the interplay of a Dirac band structure, a Zeeman field,
and $s$-wave pairing can give rise to the emergence of CTSCs,
we first consider the case where $\eta=0$. In this limit,
the two sublattices are decoupled, yielding $\mathcal{H}(\bk)=\mathcal{H}_{A}(\bk)\oplus \mathcal{H}_{B}(\bk)$,
where
\begin{eqnarray}
\mathcal{H}_{A/B}(\bk)&=&[-2t(\cos k_{x}+\cos k_{y})-\mu]\tau_{z}+M_{z}\tau_{z}s_{z}\nonumber\\
&&\pm2\lambda_{\rm so}(\sin k_{x}\tau_{z}s_{y}-\sin k_{y}s_{x})+\Delta_{s}\tau_{y}s_{y}.
\end{eqnarray}
The SOC simplifies to the Rashba form, and this particular form of the Hamiltonians $\mathcal{H}_{A}$ and
$\mathcal{H}_{B}$ is well-known for supporting CTSCs~\cite{sato2009non,sau2010}.
Despite the fact that the SOC terms have opposite signs for $\mathcal{H}_{A}$ and $\mathcal{H}_{B}$,
their Chern numbers are identical.  To facilitate a clear discussion,
we assume that $t$, $M_{z}$, and $\Delta_{s}$ are all positive, with $\Delta_{s}<M_{z}$.
By only varying the chemical potential $\mu$, we determine the dependence of the Chern number on $\mu$~\cite{Xiao2010Berry},
which is given by
\begin{eqnarray}
C_{A}=C_{B}=\left\{\begin{array}{cc}
              1, & -4t-\mu_{c}<\mu<-4t+\mu_{c}, \\
              -2, & -\mu_{c}<\mu<\mu_{c}, \\
              1, & 4t-\mu_{c}<\mu<4t+\mu_{c}, \\
              0, & \text{other cases,}
            \end{array}\right.
\end{eqnarray}
where $\mu_{c}=\sqrt{M_{z}^{2}-\Delta_{s}^{2}}$. Since $C_{A}$ equals $C_{B}$,
the Chern number of the resulting CTSCs, given by $C=C_{A}+C_{B}$, is always even. 
Remarkably, the maximum value of $C$, 
and consequently the maximum number of chiral Majorana edge states, can reach as high as $4$.

\begin{figure}[t]
\centering
\includegraphics[width=0.45\textwidth]{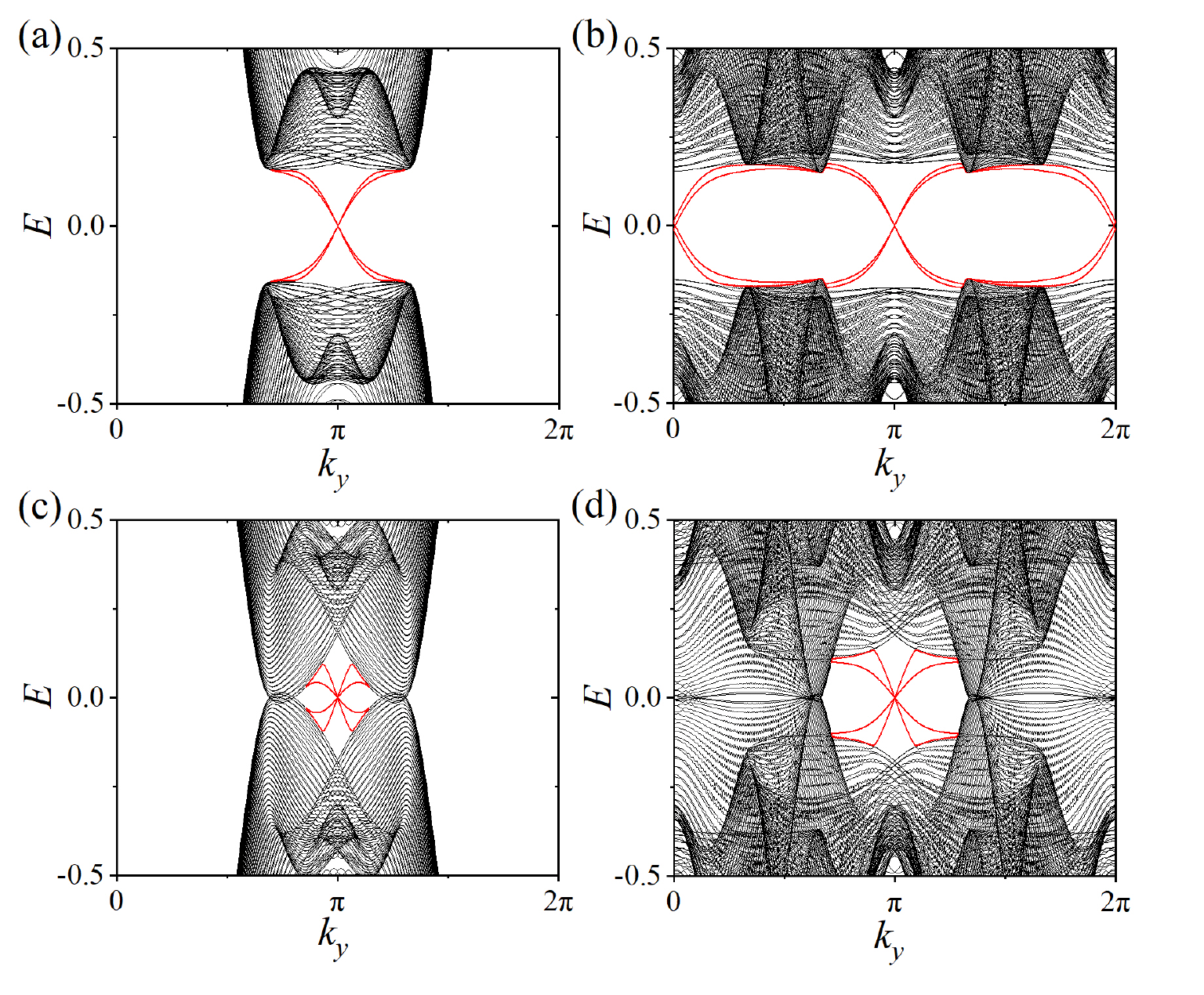}
\caption{Energy spectrum for a system with open (periodic) boundary conditions in the $x$ ($y$) direction. (a) TSC with $C=2$. (b) TSC with
$C=-4$. Figures (c) and (d) show the coexistence of BFS and chiral Majorana edge states.
The parameters $(\mu,\eta)$ are set to $(4,0.1)$ in (a), $(0,0.025)$ in (b), $(4,0.8)$ in (c) and $(0,0.2)$ in (d), respectively.
The values of the shared parameters are: $t=1$, $\lambda_{\rm so}=0.5$, $M_{z}=0.5$ and $\Delta_{s}=0.2$. 
}\label{fig3}
\end{figure}

Next, we investigate the impact of enhancing the coupling strength $\eta$ between the two sublattices.
Since a change in the Chern number necessitates the closing of the bulk energy gap, it is
evident that the CTSC can persist for a range of $\eta$ values.
Through numerical calculations, we confirm the expectation that a modest increase in $\eta$
does not induce the closure of the bulk energy gap and, consequently, exerts minimal
influence on the chiral Majorana edge states, as illustrated in Figs.\ref{fig3}(a) and \ref{fig3}(b).

Intriguingly, we find that when $\eta$ attains a critical threshold, the bulk energy gap closes.
However, further increasing $\eta$ does not reopen the gap. Instead, the middle two bands remain in contact,
forming line degeneracies. Notably, despite the absence of a bulk gap resulting in an undefined Chern number,
the chiral Majorana edge states persist if their spectral crossing occurs at $k_{y}=\pi$ [or $k_{x}=\pi$ if open (periodic) boundary
conditions are considered in the $y$ ($x$) direction, due to $C_{4z}$ rotation symmetry] of the edge Brillouin zone as illustrated in Figs.\ref{fig3}(c) and \ref{fig3}(d).
The robustness of these chiral Majorana edge states can be attributed to the vanishing of
the $\epsilon_{1}(\bk)$ term when $k_{x}$ or $k_{y}$ equals $\pi$. In simpler terms,
when the spectral crossing of these states occurs at $k_{x}=\pi$ or $k_{y}=\pi$,
increasing $\eta$ has no influence on the edge gap or the crossing at that momentum, as $\epsilon_{1}(\bk)$ identically vanishes at
those points.
In contrast, if the spectrum crossing occurs at $k_{x}=0$ or $k_{y}=0$, increasing $\eta$ 
causes both the edge gap and the spectrum crossing at those momenta to disappear, as illustrated in Fig.\ref{fig3}(d).

In two dimensions, line degeneracies in a superconductor
correspond to topologically protected BFSs. As previously mentioned,
the topological protection originates from the coexistence
of inversion symmetry and particle-hole symmetry in this system.
As first noted in Ref.~\cite{Agterberg2017BFS}, when these two symmetries
coexist, the Hamiltonian can be transformed into an antisymmetric form,
$\tilde{\mathcal{H}}(\bk)=U \mathcal{H}(\bk)U^{\dag}$, where $U$
denotes a unitary matrix. The transformed Hamiltonian obeys the relation
$\tilde{\mathcal{H}}(\bk)=-\tilde{\mathcal{H}}^{\mathbb{T}}(\bk)$, with
$\mathbb{T}$ denoting the transpose operation. The antisymmetric property allows for a
definition of Pfaffian, $P(\bk)=\text{Pf}[\tilde{\mathcal{H}}(\bk)]$.
Since $\det \mathcal{H}(\bk)=\det \tilde{\mathcal{H}}(\bk)=P^{2}(\bk)$,
the zeros of $P(\bk)$ determines the band degeneracies of $\mathcal{H}(\bk)$  at $E=0$.
Topologically protected BFSs correspond to line degeneracies across which the sign
of $P(\bk)$, denoted as $\text{sgn}[P(\bk)]$, changes from $1$ to $-1$, or vice versa.

\begin{figure}[t]
\centering
\includegraphics[width=0.45\textwidth]{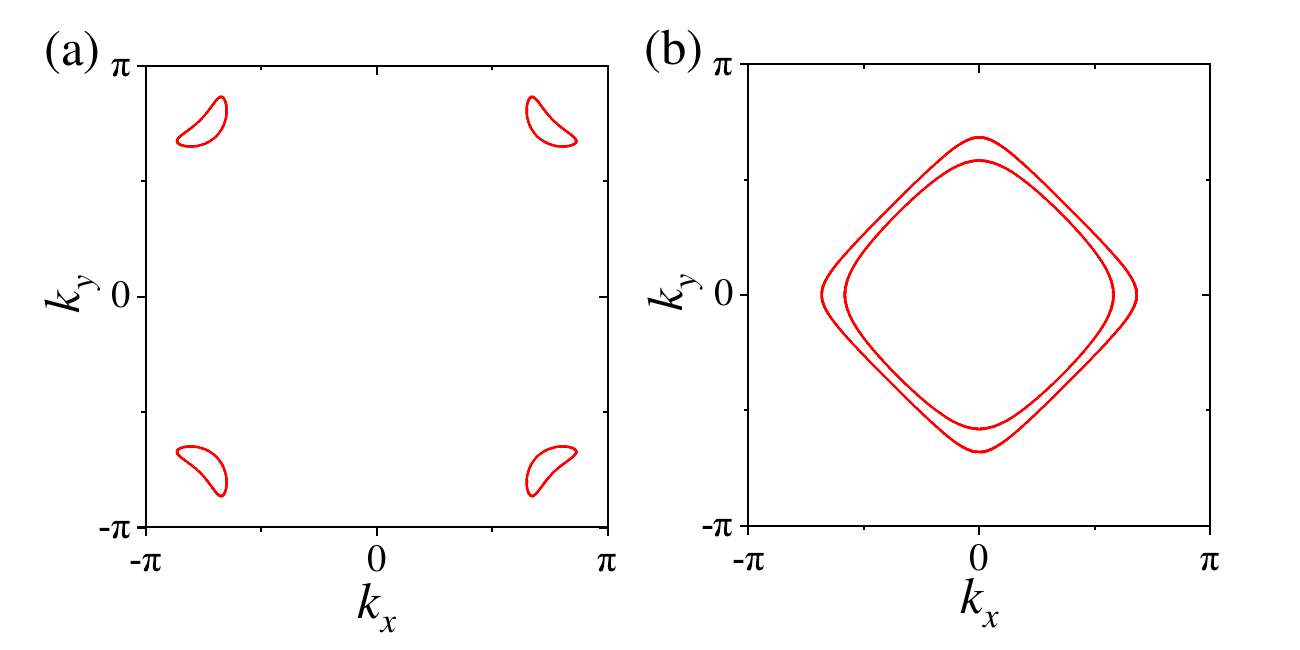}
\caption{Two representative configurations of BFSs (red lines) in the Brillouin zone. Panel (a) 
corresponds to parameters $(\mu,\eta)=(4,1)$, while panel (b) corresponds to $(\mu,\eta)=(0,0.3)$. 
The values of the shared parameters are: $t=1$, $\lambda_{\rm so}=0.5$, $M_{z}=0.5$ and $\Delta_{s}=0.2$.
}\label{fig4}
\end{figure}

For the current Hamiltonian, $P(\bk)$ is given by (see Supplemental Material (SM)~\cite{supplemental})
\begin{eqnarray}
P(\bk)&=&[\epsilon_{+}(\bk)\epsilon_{-}(\bk)+\Delta_{s}^{2}-M_{z}^{2}]^{2}+4\Delta_{s}^{2}\Lambda^{2}(\bk)\nonumber\\
&&-4\epsilon_{1}^{2}(\bk)[M_{z}^{2}-\Delta_{s}^{2}],\label{Pfaffian}
\end{eqnarray}
where $\epsilon_{\pm}(\bk)=\epsilon_{0}(\bk)-\mu\pm\sqrt{\epsilon_{1}^{2}(\bk)+\Lambda^{2}(\bk)}$. It
is evident that $\eta\neq0$ (ensuring $\epsilon_{1}(\bk)$ does not vanish identically)
and $|M_{z}|>|\Delta_{s}|$ are two necessary conditions
for $P(\bk)$ to have zeros in this system. In Fig.\ref{fig4}, we illustrate two representative configurations of BFSs
by plotting the contours where $P(\bk)$ equals zero. We have verified that the sign of $P(\bk)$
indeed changes across these BFSs, confirming their topological protection.

{\it Impact of BFSs on thermal Hall effects.---}A defining characteristic of CTSCs 
is the quantized thermal Hall effect in the low-temperature limit. 
The quantized coefficient of this effect is directly proportional to the Chern number, expressed as
$\kappa_{xy}=C\kappa_{0}$, where $\kappa_{0}=\pi k_{B}^{2}T/12\hbar$~\cite{Kasahara2018,Yokoi2021}. 
When a bulk energy gap exists, low-energy excitations are confined to the boundary, 
and the quantized effect can be derived using conformal field theory. This theory elucidates
that each branch of the chiral Majorana edge state contributes equally to the effect~\cite{read2000}. 
However, in the presence of BFSs, low-energy excitations also 
emerge in the bulk. These low-energy Bogoliubov quasiparticles near the BFSs experience a nonzero Berry curvature, resulting in a nonzero contribution to the thermal Hall effect that is generally non-quantized~\cite{Wang2021Berry}. 
To investigate the influence of BFSs on the thermal Hall effect, we utilize the formula for the thermal Hall coefficient 
developed in Refs.~\cite{Qin2011THE,Sumiyoshi2013THE}, which is given by 
\begin{eqnarray}
\kappa_{xy}=-\frac{k_{B}^{2}T}{2\hbar}\int_{-\infty}^{+\infty}dE\frac{E^{2}}{(k_{B}T)^{2}}\sigma(E)f'(E),
\end{eqnarray}
where $f(E)=1/[\exp(E/k_{B}T)+1]$ is the Fermi-Dirac distribution function, $k_{B}$ is the Boltzmann constant (we set $k_{B}=1$), 
$T$ is the temperature, and 
\begin{eqnarray}
\sigma(E)=\sum_{n}\int_{E_{n}(\bk)<E}\frac{d^{2}k}{(2\pi)^{2}}\Omega^{(n)}(\bk).
\end{eqnarray}
Here $\Omega^{(n)}(\bk)=-2\text{Im}\langle\partial_{k_{x}}u_{n}(\bk)|\partial_{k_{y}}u_{n}(\bk)\rangle$ 
denotes the Berry curvature of the $n$th band.

\begin{figure}[t]
\centering
\includegraphics[width=0.45\textwidth]{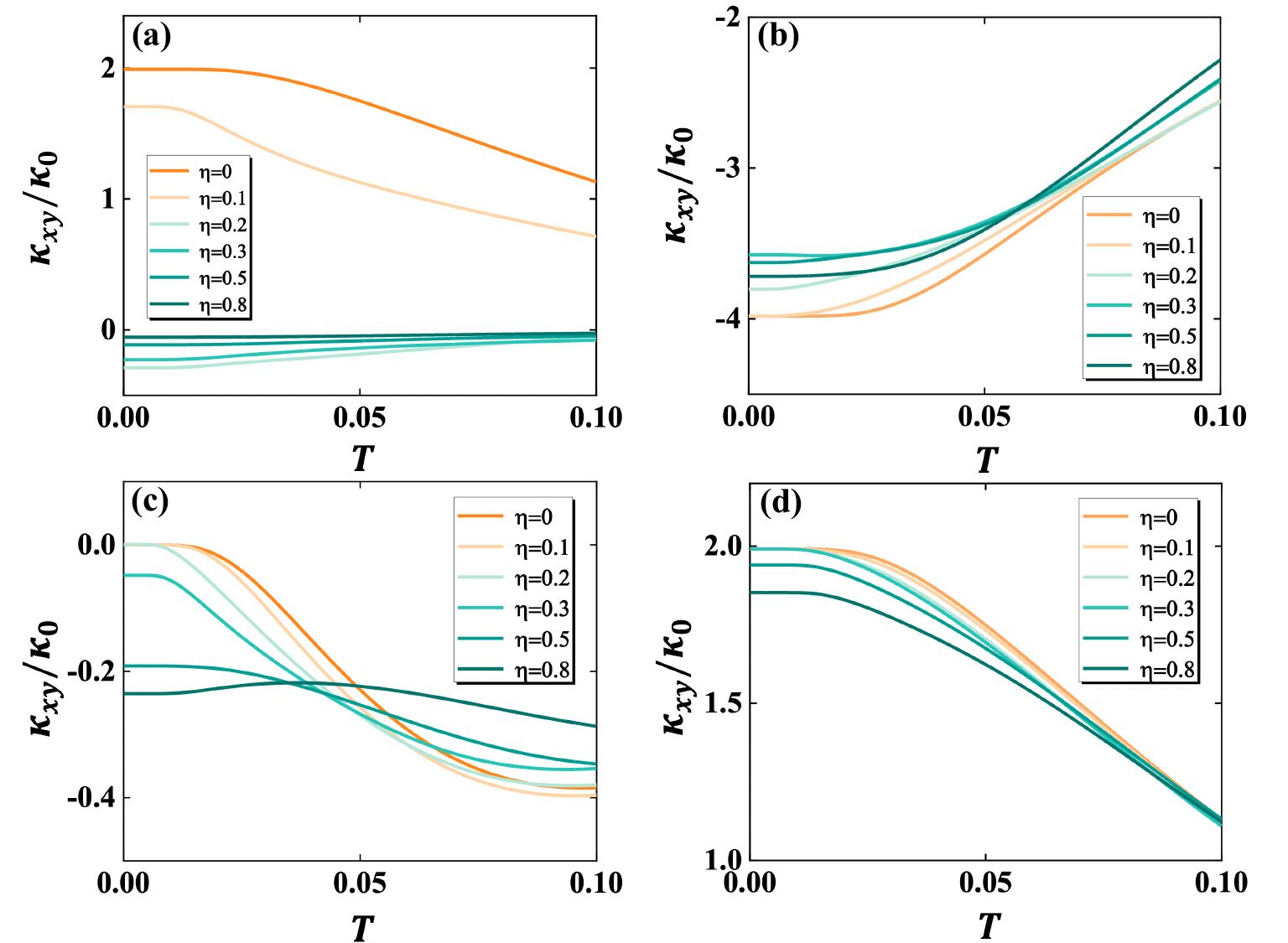}
\caption{Evolution of the thermal Hall coefficient under the variation of $\eta$. 
The values of the shared parameters are: $t=1$, $\lambda_{\rm so}=0.5$, $M_{z}=0.5$ and $\Delta_{s}=0.2$.  
(a) $\mu=-4$, (b) $\mu=0$, (c) $\mu=1$, (d) $\mu=4$. The critical value of $\eta$ for the onset of BFSs is approximately equal to
0.11 in (a), 0.19 in (b), 0.27 in (c) and 0.72 in (d).  
The Chern number before the onset of BFSs is equal to $2$ in (a), $-4$ in (b), $0$ in (c) and $2$ in (d).
}\label{fig5}
\end{figure}

In Fig.\ref{fig5}, we present the evolution of $\kappa_{xy}$ under the variation of $\eta$. {\zb 
Although the pairing amplitude depends on temperature, here we neglect this dependence. A 
detailed discussion of the temperature-dependent pairing case is provided in the SM~\cite{supplemental}. We emphasize 
that this approximation is rather accurate across a wide range of low temperatures, which is the primary focus of our study.} 

Specially, the Chern number in Figs.\ref{fig5}(a), \ref{fig5}(b), \ref{fig5}(c) and \ref{fig5}(d) is $2$, $-4$, $0$, and $2$, respectively, before $\eta$ surpasses the critical value, denoted as $\eta_{c}$, where BFSs emerge. Notably, $\eta_{c}$ depends on other parameters, as can be inferred from Eq.(\ref{Pfaffian}). The results depicted in Fig.\ref{fig5} demonstrate quantized behaviors in the low-temperature limit, provided that the bulk energy gap remains open. As anticipated, once $\eta$ exceeds $\eta_{c}$, the quantized behavior is disrupted due to the presence of BFSs. Intriguingly, we observe that the change in $\kappa_{xy}$ caused by BFSs is most pronounced in Fig.\ref{fig5}(a). For the largest value of $\eta$ considered, 
$\kappa_{xy}$ in Fig.\ref{fig5}(a) deviates significantly from its original quantized value, while 
the results in Fig.\ref{fig5}(b) and Fig.\ref{fig5}(d) show much smaller derivations. 
The remarkable change observed in Fig.\ref{fig5}(a) can 
be attributed to the fact that the spectrum crossing of chiral Majorana edge states occurs at $k_{x}=0$ or $k_{y}=0$
for this particular case. As discussed earlier, 
the presence of BFSs causes these chiral Majorana edge states to disappear.

{\it Discussions and conclusions.---}We demonstrated that applying a Zeeman 
field perpendicular to the spin plane of the inversion-symmetric SOC 
can transform the nonsymmorphic DSM into a Chern insulator. Furthermore, 
we showed that incorporating $s$-wave superconductivity into this system 
can lead to the realization of fully-gapped CTSCs with large Chern numbers, 
as well as gapless superconducting phases where BFSs coexist with chiral Majorana edge states. 
Additionally, we found that the presence of BFSs can have a significant impact on the thermal Hall effect. 

{\zb In addition to the thermal Hall effect, the  
chiral Majorana edge states and BFSs can be experimentally probed using a scanning tunneling microscope to 
measure the density of states at the edges and in the bulk, respectively. 
The chiral Majorana edge states are characterized by a finite and constant 
density of states localized around the edges, while the BFSs contribute a finite 
density of states at zero energy across the system. Interestingly, we observe that 
in this system, the density of states develops a zero-bias peak at the onset of BFSs. 
This robust and distinctive signature is expected to be readily measurable~\cite{supplemental}. 
Regarding the experimental realization of our proposed setup, we highlight that 
a variety of 2D DSMs have been theoretically predicted. These include
black phosphorus~\cite{Kim2017DSM}, FeSe~\cite{Young2017DSM}, 
X$_3$SiTe$_6$ ($X$ = Ta, Nb)~\cite{Li2018NSDSM,Sato2018NSDSM}, 
SbSSn~\cite{Jin2020DSM}, HfGe$_{0.92}$Te~\cite{Chen2022DSM}, and others~\cite{Meng2022DSM},  
several of which have already been experimentally confirmed.
Moreover, suitable superconductor substrates are abundant, as only conventional 
$s$-wave pairing is required. These factors ensure that our predictions are well within 
the realm of experimental feasibility.}

In conclusion, our study underscores the rich interplay between Zeeman fields, superconductivity, 
and nonsymmorphic Dirac band structures, which can give rise to a diverse array of 
phases exhibiting intriguing properties.

{\it Acknowledgements.---}This work is supported by the National Natural Science Foundation of China (Grant No.12174455),
Natural Science Foundation of Guangdong Province
(Grant No. 2021B1515020026), and Guangdong
Basic and Applied Basic Research Foundation (Grant No.
2023B1515040023).

\bibliography{dirac}

\end{document}